\documentclass[prl,aps,floatfix,twocolumn,linenumbers]{revtex4-1}
\usepackage{endnotes}
\usepackage{latexsym,amssymb}
\usepackage{amsmath, amsthm, amssymb}
\usepackage{graphicx}

\begin{document}
\widetext
\leftline{To be submitted to PRL}



\title{Kinetics of microribbon formation in a simplified model of amelogenin biomacromolecules}

\author{Wei Li}
\email{wel208@lehigh.edu}
\affiliation{Department of Physics, Lehigh University, Bethlehem, PA 18015}

\author{Ya Liu}
\affiliation{Department of Physics, Lehigh University, Bethlehem, PA 18015}

\author{Toni Perez}
\affiliation{Department of Physics, Lehigh University, Bethlehem, PA 18015}

\author{J. D. Gunton}
\affiliation{Department of Physics, Lehigh University, Bethlehem, PA 18015}

\author{C. M. Sorensen}
\affiliation{Department of Physics, Kansas State University, Manhattan, KS 66503}

\author{A. Chakrabarti}
\affiliation{Department of Physics, Kansas State University, Manhattan, KS 66503}

\date{\today}

\begin{abstract} 
We show that the kinetics of microribbon formation of amelogenin molecules is well described by a combination of translational and rotational diffusion of a simplified anisotropic bipolar model consisting of hydrophobic spherical colloid particles and a point charge located on each particle surface. The colloid particles interact via a standard depletion attraction while the point charges interact through a screened Coulomb repulsion. We study the kinetics via a Brownian dynamics simulation of both translational and rotational motions and show that the anisotropy brought in by the charge dramatically affects the kinetic pathway of cluster formation and our simple model captures the main features of the experimental observations.
\end{abstract}

\maketitle


The self-assembly of colloidal particles and globular proteins with inhomogenous properties into various desired structures is an important area of current research, in which the inhomogeneity (such as an amphiphilic group and a residual group) plays a crucial role in the kinetic pathway of cluster formation and eventually impacts on the stabilized crystal structures. Examples include Janus particles and DNA-coated colloidal particles that lead to the next generation of building blocks of new materials and have potential applications in fabricating phontonic crystals, targeted drug delivery, and electronic equipment \cite{Qian11,Flavio11,Hong08,Mirkin10}. Another important example is amelogenin, the chief hydrophobic protein in enamel matrix with a hydrophilic 25-amino acid C terminus. Amelogenin is involved in the mineral deposition and is responsible for the major structural process during enamel biomineralization. A recent experiment shows that the hydrophilic C-terminus is essential for the self-assembly of amelogenin into microribbons, which may also be relevant to the elongated and oriented growth of apatite crystals during biomineralization \cite{Du05,Ame07}. The experiment also revealed that the microribbon is formed by a hierarchical organization of amelogenin molecules into a chain of nanospheres.

In this article we present a coarse-grained model of amelogenin and explain how the microribbon forms through self-assembly. The amelogenin molecule is hydrophobic with a charged hydrophilic tail. Our model describes this in a simplified way, representing the monomer as a spherical molecule, with the charged hydrophilic tail replaced by a single tethered point charge located on the surface of the molecule. Using Brownian dynamics simulations, we investigate the static and dynamic properties of the self-assembly process. We show that the anisotropy brought by the charge dramatically affects the kinetic pathway of cluster formation and our simple model captures the main features of the experimental observations in the early stages.

In the experimental studies of amelogenin assembly process, one typically adds salt and a precipitant such as polyethylene glycol (PEG). Although PEG-protein interactions are quite complicated \cite{prl06}, we model this as a depletion interaction that we take to be given by the Asakura-Oosawa potential plus a repulsive hard-core-like interaction, depending on the center-center distance between spherical particles, $U_{c}(r_{ij}^{c})=U_{AO}(r_{ij}^{c})+U_{hc}(r_{ij}^{c})$,
where
\begin{eqnarray}
\frac{U_{AO}(r_{ij}^{c})}{kT}=\left\{\begin{array}{rcl}
\phi_p(\frac{r_{c}}{\xi})^3[\frac{3r_{ij}^{c}}{2r_{c}}-\frac{1}{2}(\frac{r_{ij}^{c}}{r_{c}})^3 - 1], & r_{ij}^{c}<r_{c}
\\ 0, & r_{ij}^{c}>r_{c}
\end{array}\right.
\end{eqnarray}
The cut-off range $r_{c}\equiv1+\xi$, where $\xi$ is the size-ratio between a PEG coil and a colloidal amelogenin particle that controls the range of the depletion interaction, and $\phi_{p}$ is the value of the PEG volume fraction that controls the strength of the interaction described by the absolute value of the minimum potential depth $U_{m}\equiv|U_{min}|$ \cite{AO54,Vrij76}. The hard core potential is given by
\begin{eqnarray}
\frac{U_{hc}(r_{ij}^{c})}{kT} = (r_{ij}^{c})^{-\alpha}.
\label{xibulk}
\end{eqnarray}
We set $\alpha= 36$, since the values of $\alpha < 36$ have been reported to lead to anomalies when a mimic of the hard-core potential is required in the potential \cite{A36}. The point charges interact with each other through a screened Coulomb potential,
\begin{eqnarray}
U_{p}(r_{ij}^{p})=\frac{\varepsilon}{r_{ij}^{p}}\exp\bigg(-\frac{r_{ij}^{p}}{\lambda_{D}}\bigg),
\label{xibulk}
\end{eqnarray}
in which magnitude controlled by $\varepsilon$ and range controlled by Debye screening length $\lambda_{D}$, are varied in the simulations. We find that the early morphology of the self-assembly is more sensitive to the size of $\lambda_{D}$ rather than $\varepsilon$. These Coulomb charges exert torques on adjacent molecules and hence produce a rotational motion of the molecules that is included in our Brownian equations of motion, read as:
\begin{eqnarray}
m\ddot{\vec{r}}_{i}=-\vec{\nabla}(U_{i}^{c}+U_{i}^{p})-\Gamma_{t}\dot{\vec{r}}_{i}+\vec{W}_{i}(t),
\label{xibulk} 
\end{eqnarray}
\begin{eqnarray}
I\dot{\vec{\omega}}_{i}=\vec{\tau}_{i}-\Gamma_{r}\vec{\omega}_{i}+\vec{W}'_{i}(t),
\label{xibulk}
\end{eqnarray}
where $m$, $I$, $\vec{r}_{i}$, $\vec{\omega}_{i}$, $\vec{\tau}_{i}$ are the mass, moment of inertia, position vector, angular velocity, and torque, respectively, of the ith colloidal particle. The mass of the point charge is ignored in this model. $\Gamma_{t}$ ($\Gamma_{r}$) is the translational (rotational) friction coefficient. $\vec{W}_{i}$ and $\vec{W}'_{i}$ are the random forces and torques acting on the ith colloidal particle respectively, which satisfy a fluctuation-dissipation relation $\langle\vec{W}_{i}(t)\cdot\vec{W}_{j}(t')\rangle=6kT\Gamma_{t}\delta_{ij}\delta(t-t')$, $\langle\vec{W}'_{i}(t)\cdot\vec{W}'_{j}(t')\rangle=6kT\Gamma_{r}\delta_{ij}\delta(t-t')$ \cite{VG82}. We choose $\Gamma_{t}=0.5$, $\Gamma_{r}=0.167$ and the time step $\triangle t=0.005$ in reduced time units of $\sigma(m/kT)^{1/2}$ with $m=1$ \cite{Li11}. The other parameters are chosen according to the range of values of the experimental data displayed in Du et al's paper \cite{Du05}. In particular, here the results shown are for the values $\xi$=0.1, $U_{m}=6kT$, $\lambda_{D}=0.4$ (all length scales are measured in units of monomer diameter and energies are scaled by $\varepsilon$). We consider a small volume fraction of the molecules, typically of the order of $f=0.01$ or $0.02$. Periodic boundary conditions are enforced to minimize wall effects. All simulations start from a random initial monomer conformation and the results for the kinetics are averaged over several (5-10) runs.

We show typical configurations in the early stage of microribbon formation in Fig. 1. We see that initially the monomers form oligomers and the oligomers
self-assemble into larger aggregates contained (approximately) within nanospheres, then nanospheres associate together into a chain structure, quite similar to the
process of amelogenin self-assembly (see Fig. 3 in reference \cite{Du05}).
\begin{figure}[htbp] \includegraphics[width=0.8\linewidth]{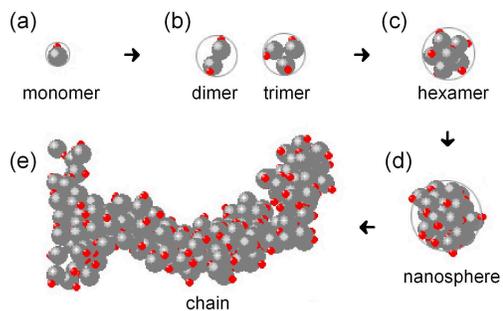}
\caption{\label{fig:epsart}(color online). (a) The model of amelogenin molecule consists of a spherical colloid particle and a point charge located on its surface that preserves a bipolar nature. (b) and (c) Oligomerization of amelogenin molecules occurs by means of hydrophobic interactions and modification due to Coulomb repulsions. (d) Nanosphere structures are formed through aggregation of monomers and oligomers. (e) Further association of nanospheres results in larger assemblies among which chains are formed. }
\end{figure}

The Coulomb forces cause the majority of the charges to point outwards inside the nanospheres, as shown in Fig. 2. This again is reminiscent of the situation with the hydrophilic tails of amelogenin. To characterize the orientational distribution of point charges within a nanosphere, we introduce a parameter $\theta$, defined as $\theta = \cos^{-1}\bigg(\frac{\vec{u}_p \cdot \vec{u}_c}{|\vec{u}_p||\vec{u}_c|}\bigg)$, where $\vec{u}_p$ is the position vector of a point charge referenced to the center of its host colloid and $\vec{u}_c$ the position vector of a colloid referenced to the center of mass of this nanosphere. Therefore $\theta < 90^{\circ}$ if the charge points outwards and $\theta > 90^{\circ}$ if it points inwards. Due to fluctuations and our choice of short repulsion interaction range, the distribution of $\theta$ is spread around a small nonzero peak position, as shown in Fig. 2, for the configuration in inset. Subsequent aggregation forms via necks between nanospheres, leading to larger structures that eventually form (flexible) microribbons.
\begin{figure}[htbp] \includegraphics[width=0.8\linewidth]{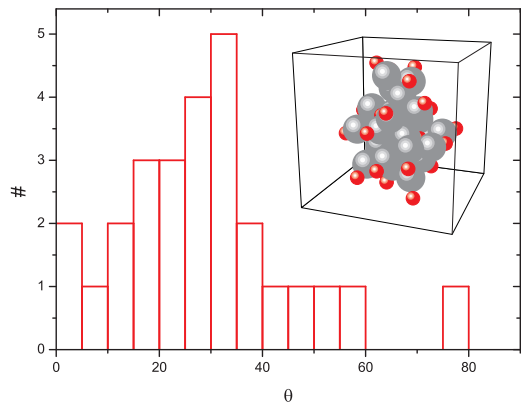}
\caption{\label{fig:epsart}(color online). Distribution of $\theta$; $\theta$ shows whether the point charge orients outwards ($\theta<90^{\circ}$) or inwards ($\theta>90^{\circ}$). Inset: Nanosphere structure, large (grey) spheres indicate amelogenin molecules while small (red) ones for point charges. }
\end{figure}

We characterize the morphology of the clusters in terms of their fractal dimension $D_{f}$. The q-dependence of $S(q,t)$ shown in Fig. 3(a) is given by a slope of -2
on a log-log plot over a reasonable range of q. This value of $D_{f}$ is larger than the typical diffusion-limited-cluster-cluster aggregation (DLCA) value of 1.8
as the ribbon-like clusters observed here have significant short range ordering \cite{Porod}. For this reason we speculate that the repulsive Coulomb force causes the surface particles to reorganize, which reduces the surface roughness of our model, in contrast to the case without the Coulomb interactions (see reference \cite{Li11}). We find that a peak develops in the structure factor whose magnitude increases with time. Correspondingly, the peak position decreases as a function of wave number with increasing time; these features are typical of spinodal decomposition \cite{Gunton}. However, we note that for deep quenches that lead to cluster growth considered in this model, the system is controlled by two characteristic lengths that evolve differently in time and an apparent scaling for the structure factor can only be observed over some period of time when these two characteristic length scales become comparable to each other \cite{Amit04,PRE98}.
\begin{figure}[htbp] \includegraphics[width=0.8\linewidth]{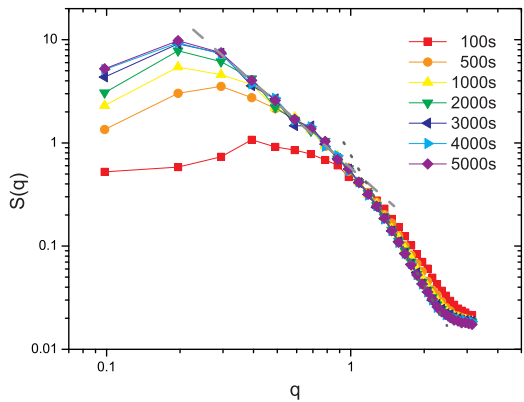}
\caption{\label{fig:epsart} (color online). Log-log plot of structure factor at several times. The dashed line indicates fractal clusters with $S(q)\propto q^{-D_{f}}$, where $D_{f}=2.0$, while the dotted line indicates the Porod regime $S(q)\propto q^{-(d+1)}$, $d=3$. }
\end{figure}

The kinetics of the cluster growth process in this simple model for amelogenin self-assembly is consistent with a cluster-cluster aggregation mechanism, as shown in
Fig. 4. The number of clusters decreases inversely with time (i.e. the kinetic exponent $z=1$) and the radius of gyration increases as a power law with an exponent
of $n=0.5$ \cite{KF,Amit}. The relation between $z$ and $n$ involves the fractal dimension in the following way: $n=z/D_{f}$. Thus the kinetic exponents are consistent with a fractal dimension of $D_{f}=2$ as well.
\begin{figure}[htbp] \includegraphics[width=0.8\linewidth]{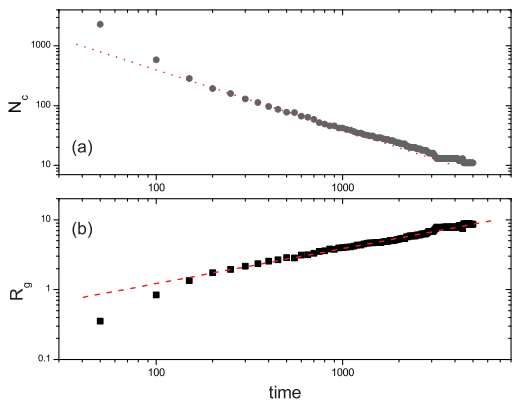}
\caption{\label{fig:epsart} (color online). Plot of (a) number of clusters, $N_{c}$ and (b) radius of gyration, $R_{g}$ as a function of time in a log-log plot. The dotted line in (a) has a slope of $-1$; the dashed line in (b) has a slope of $0.50$.}
\end{figure}

In order to check the stability of the clusters produced in straight simulation from a quench ($U_{m}=6kT$) into the two phase gas-solid region, we ``heat" the system formed after $5000$ steps from the initial quench $U_{m}=6kT$ to $U_{m}=4kT$ and run the simulation for another $5000$ time steps. This heating corresponds to the change in PEG concentration as it is done in the experiment. A typical run is shown in Fig. 5. As shown in the figure, we find that the microribbons are quite robust under such a change of conditions, as found experimentally for a wide range of conditions (including a change in the PEG concentration). We have also checked the structure factor behavior for the heating process and found that the structure factor curve has the same exponent of -2 as at the cooler temperature (see Fig. 3). This is consistent with the robustness of the major cluster morphology.
\begin{figure}[htbp] \includegraphics[width=0.8\linewidth]{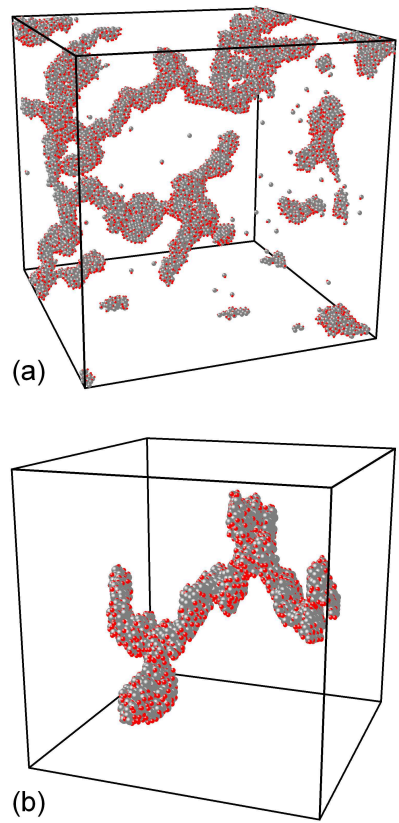}
\caption{\label{fig:epsart} (color online). (a) Morphology of entire system ($\xi=0.1$, $f=0.02$) from another $5000$ time steps shallow quench of $U_{m}=4kT$ after first quench into $U_{m}=6kT$, $5000$ time steps. (periodic boundary condition are enforced). (b) One of the clusters inside (a).}
\end{figure}

We characterize these microribbons by computing three eigenvalues of the gyration tensor, denoted as $\lambda_{1}^{2}$, $\lambda_{2}^{2}$, and $\lambda_{3}^{2}$ in descending order \cite{gt85}. We calculate the ratios among these three principal values, which characterize the relative colloid (or charge) particle distribution. A typical result is given in Table 1, for the colloid particles and point charges separately, based on the configuration in Fig. 5(b). These ratios of eigenvalues demonstrate the anisotropy of these clusters, suggesting that the cluster is ribbon-like. The difference of the ratios between colloidal particles and point charges may indicate that under electrical repulsion, the rotation of monomers inside the cluster leads to the small reduction of anisotropy in the point charges' spatial distribution.
\begin{table}
\caption{\label{tab:table1} Typical ratios of eigenvalues of gyration tensor}
\begin{ruledtabular}
\begin{tabular}{lcr} &\mbox{$\lambda_{1}^{2}/\lambda_{3}^{2} $} &\mbox{$ \lambda_{2}^{2}/\lambda_{3}^{2} $} \\
\hline
Colloid particles & 9.57 & 9.03\\
Point charges &	9.25 & 8.74 \\
\end{tabular}
\end{ruledtabular}
\end{table}

From the gyration tensor we can determine the typical length scales of these clusters by taking square roots of the eigenvalues. To illustrate this, we use the colloid configuration shown in Fig. 5(b), which gives $18.27$, $17.75$, and $5.91$ for the three length scales. Since the chain structure is the association of several nanospheres (as shown in Fig. 1), the typical diameter of the nanosphere in this configuration is at the most $5.91$. We can also use this estimate of the diameter to obtain an estimate of the minimum wavenumber that corresponds in Fig. 3 to the crossover to the Porod regime (corresponding to compact clusters on a local scale). This yields a value of $q_{min} \simeq 2\pi/5.91 = 1.06$, which is consistent with the behavior shown in Fig. 3(b). Since the hydrodynamic radius $R_{H}$ of the amelogenin macromolecule is about $2.2$ nm, the $R_{H}$ of the nanosphere for the configuration in Fig. 5(b) is $2.2\times5.91 = 13.00$ nm; this lies in the range of values of the radii of typical nanospheres, $10\sim 25$ nm, found in the experiment (cf. Fig. 3, reference \cite{Du05}).

In summary, we have developed an anisotropic, bipolar model for the hierarchical self-assembly of amelogenin molecules and have carried out Brownian dynamics simulations of the self-assembly process. Simulations show a hierarchical self-assembly process where the molecules aggregate to form dimers, hexamers to nanospheres, and the assembly of the nanospheres then lead to the formation of microribbons in agreement with experimental findings. The relative strengths of the interaction parameters can lead to a phase diagram where the nanospheres are stable against microribbon formation. Thus a control over the morphology of the clusters can be achieved. Such a study is currently underway.

This work was supported by grants from the Mathers Foundation and the National Science Foundation (Grant DMR-0702890). Work at Kansas State University was supported by NSF NIRT grant CTS0609318.


\begin{thebibliography}{99}

  \bibitem{Qian11}
  Qian Chen, Sung Chul Bae, and Steve Granick, Nature {\bf 469}, 381, (2011).

  \bibitem{Flavio11}
  Flavio Romano and Francesco Sciortino, Nature material {\bf 10}, 171, (2011).

  \bibitem{Hong08}
  L. Hong, A. Cacciuto, E. Luijten, and S. Granick, Langmuir {\bf 24}, 621-625 (2008).

  \bibitem{Mirkin10}
  Chad A. Mirkin, MRS Bulletin {\bf 35}, 532-539 (2010).

  \bibitem{Ame07}
  S. Gajjeraman, K. Narayanan, J. Hao, C. Qin and A. George, J. Biological chemistry {\bf 282}, 1193 (2007).

  \bibitem{Du05}
  C. Du, G. Falini, S. Fermani, C. Abbott, J. Moradian-Oldak, Science {\bf 307}, 1450 (2005).

  \bibitem{prl06}
  J. Bloustine, T. Virmani, G. M. Thurston, and S. Fraden, Phys. Rev. Lett. {\bf 96}, 087803 (2006).

  \bibitem{AO54}
  S. Asakura and F. Oosawa, J. Chem. Phys. {\bf 22}, 1255 (1954).

  \bibitem{Vrij76}
  A. Vrij, Pure Appl. Chem. {\bf 48}, 471 (1976).

  \bibitem{A36}
  K. G. Soga, J. R. Melrose, and R. C. Ball, J. Chem. Phys. {\bf 108}, 6026 (1998); K. G. Soga, J. R. Melrose, and R. C. Ball, J. Chem. Phys. {\bf 110}, 2280
   (1999).

  \bibitem{VG82}
  W. F. Van Gunsteren, and H. J. C. Berendsen, Mol. Phys. {\bf 45}, 637 (1982).

  \bibitem{Li11}
  Wei Li, J. D. Gunton, Siddique J. Khan, J. K. Schoelz, and A. Chakrabarti, J. Chem.  Phys. {\bf 134}, 024902 (2011).

  \bibitem{Porod}
  G. Porod, Zeit. Kolloid, {\bf 124}, 53 (1951). See also, O. Glatter and O. Kratky. Small-angle X-ray Scattering, (Academic Press, London 1982).

  \bibitem{Gunton}
  J. D. Gunton, M. San Miguel, and P. S. Sahni, in Phase Transitions and Critical Phenomena, edited by C. Domb and J. L. Lebowitz (Academic, London, 1983), Vol. 8.

  \bibitem{Amit04}
  J. J. Cerda, T. Sintes, C. M. Sorensen, and A. Chakrabarti, Phys. Rev. E {\bf 70}, 051405 (2004).

  \bibitem{PRE98}
  H. Huang, C. Oh, and, C.M. Sorensen, Phys. Rev. E {\bf 57}, 875 (1998).

  \bibitem{KF}
  S. K. Friedlander. Smoke, Dust, and Haze: Fundamentals of Aerosol Behavior (Wiley, New York, 1977).

  \bibitem{Amit}
  S. J. Khan, C.M. Sorensen, and A. Chakrabarti, J. Chem. Phys. {\bf 131}, 194908 (2009).

  \bibitem{gt85}
  Doros N. Theodorou and Ulrich W. Suter, Macromolecules, {\bf 18}, 1206 (1985).

\end{thebibliography}
\end{document}